\documentclass[review]{elsarticle}

\usepackage{lineno,hyperref}

\journal{Computer Speech and Language}

\usepackage{numcompress,amsmath,graphicx,epsfig,multicol,url}
\usepackage{multirow}
\usepackage{booktabs}
\usepackage{dcolumn}
\usepackage{tcolorbox}
\usepackage{pdflscape}
\usepackage{blindtext, rotating}
\usepackage{changepage}
\usepackage{verbatim}
\usepackage{makecell}
\usepackage{xcolor}
\usepackage{outlines}
\usepackage{siunitx}

\usepackage{rotating}

\usepackage{multirow}
\usepackage{hyperref}

\usepackage{tikz}
\usetikzlibrary{positioning}

\newcommand*\XX{\mathbf{X}}
\newcommand*\YY{\mathbf{Y}}
\newcommand*\ZZ{\mathbf{Z}}
\newcommand*\xx{\mathbf{x}}
\newcommand*\yy{\mathbf{y}}

\newcommand*\VV{\mathbf{V}}
\DeclareMathOperator{\tr}{tr}

\newcommand\myatopthree[3]{\left[\substack{#1 \\ #2 \\ #3}\right]} 

\makeatletter
\newcommand{\printfnsymbol}[1]{%
  \textsuperscript{\@fnsymbol{#1}}%
}
\makeatother

\usepackage[framemethod=tikz]{mdframed}
\usepackage[noend]{algorithmic}

\title{Bayesian HMM clustering of x-vector sequences (VBx) in speaker diarization: theory, implementation and analysis on standard tasks}
\tnotetext[t1]{The work was supported by Czech National Science Foundation (GACR) project ``NEUREM3'' No. 19-26934X, and European Union’s Horizon 2020 project No. 833635 ROXANNE.}

\begin{document}

\begin{frontmatter}


\author[1]{Federico Landini\corref{cor1}%
\fnref{fn1}}
\ead{landini@fit.vutbr.cz}
\author[2]{J\'an Profant\fnref{fn1}}
\ead{jan.profant@phonexia.com}
\author[1]{Mireia Diez}
\ead{mireia@fit.vutbr.cz}
\author[1]{Luk\'a\v{s} Burget}
\ead{burget@fit.vutbr.cz}

\cortext[cor1]{Corresponding author}
\fntext[fn1]{Equal contribution}
\address[1]{Brno University of Technology, Faculty of Information Technology, Speech@FIT, Czechia}
\address[2]{Phonexia, Czechia}


\begin{abstract}

The recently proposed VBx diarization method uses a Bayesian hidden Markov model to find speaker clusters in a sequence of x-vectors.
In this work we perform an extensive comparison of performance of the VBx diarization with other approaches in the literature and we show that VBx achieves superior performance on three of the most popular datasets for evaluating diarization: CALLHOME, AMI and DIHARDII datasets.
Further, we present for the first time the derivation and update formulae for the VBx model, focusing on the efficiency and simplicity of this model as compared to the previous and more complex BHMM model working on frame-by-frame standard Cepstral features.
Together with this publication, we release the recipe for training the x-vector extractors used in our experiments on both wide and narrowband data, and the VBx recipes that attain state-of-the-art performance on all three datasets.
Besides, we point out the lack of a standardized evaluation protocol for AMI dataset and we propose a new protocol for both Beamformed and Mix-Headset audios based on the official AMI partitions and transcriptions. 
\end{abstract}



\end{frontmatter}



\section{Introduction}
\label{sec:intro}

In recent years, speaker diarization works are proliferating. This is due to two main factors: first, new datasets and challenges are providing new benchmarks that bring the interest of the community and foster \textit{healthy} competition. The most relevant examples are the DIHARD series addressing diarization in a wide variety of challenging domains \cite{DIHARDCorpora,DIHARD19}, CHIME6 \cite{watanabe2020chime6} providing a very demanding multi-channel benchmark for diarization, or the recent VoxConverse \cite{chung2020spot} exploring diarization on several kinds of YouTube videos. Second, the success of the new end-to-end paradigm for speaker recognition is starting to be adopted for diarization tasks. Unlike standard diarization approaches which normally deal with diarization tasks using oracle voice activity detection (VAD), end-to-end diarization systems deal also with the VAD task. The end-to-end approaches have the advantage of being able to cope with overlapped speech \cite{fujita2020endtoend,huang2020speaker,Medennikov_2020TSVAD}. Even if these methods are still limited, e.g. they are restricted to scenarios with a fixed number of speakers, and mainly tested on artificially created short recordings \cite{fujita2020endtoend,kinoshita2020integrating}, and mostly do not achieve state-of-the-art results \cite{horiguchi2020endtoend}, this research line is very promising and indeed prolific.

However, due to the several difficulties that end-to-end approaches still have to overcome for diarization tasks, in recent speaker diarization evaluations the best performing systems are based on a more conventional approach, the clustering of x-vectors \cite{Sell2018b,landini2020but,xiao2020microsoft}.

In this paper, we show how our current VBx diarization approach, which clusters x-vectors using a Bayesian hidden Markov model (BHMM) \cite{diez2020optimizing}, combined with a ResNet101 x-vector extractor \cite{he2016deep} achieves superior results on CALLHOME \cite{Callhome}, AMI \cite{carletta2005ami} and DIHARDII \cite{DIHARD19} datasets.

Besides establishing new baselines for these representative datasets, we perform a thorough analysis comparing our results to the best numbers found in the literature.
In the case of AMI dataset, this proved to be a challenging task. Most works published on AMI data choose their own partition, references and audio setup for evaluation, making the comparison between works very complicated. 
We identified works that were presenting superior performance and reproducible setups and we evaluated our system in their respective setups.
In this paper, we further provide our own evaluation protocol, which comprises lists for train/dev/eval partition, references and audios. Our setup is based on the official partition of AMI corpus. 
We believe that this setup could serve as a new standard facilitating a fair comparison of diarization systems on AMI corpus.

The VBx diarization approach has been presented before \cite{DiezInter19}, but the paper did not provide any derivation of update formulae, as it was introduced merely as a special case and simplification of its \textit{big-brother} BHMM with eigen-voice priors \cite{DiezTASLP20}.
The more complex BHMM model from \cite{DiezTASLP20} operates directly on frame-by-frame standard Cepstral features. It is based on a Bayesian HMM where states correspond to speakers and transitions correspond to speaker turns. 
To robustly model speaker distributions it uses an i-vector like model  \cite{dehakivectran,KennyBayDiar}: the distribution of each speaker is represented by a Gaussian mixture model (GMM). Such GMM is constrained to live in a low-dimensional eigen-voice subspace and each speaker can be therefore robustly represented by a fixed-length i-vector like latent variable.

The VBx used in this paper is based on a similar BHMM model and a similar idea for modeling speaker distributions. However, it is used for directly clustering x-vectors, which allows to use a much simpler probabilistic linear discriminant analysis (PLDA) based model for modeling speaker distributions.
The model is essentially the same as the one in \cite{DiezTASLP20} but using only a single Gaussian component to model speaker distributions.
In fact, when VBx was introduced in \cite{DiezInter19}, it was suggested that the same model and the same inference from the BHMM model working on a frame-basis \cite{DiezTASLP20} could be reused just by replacing the GMMs that modeled the speaker-specific distributions by single Gaussians.
However, naively re-implementing the algorithm used in \cite{DiezTASLP20} is not effective as the design changes made to obtain the VBx model lead to significant simplification of the inference formulae and derivations.
Therefore, in this paper, we present the same derivations and update formulae as in \cite{DiezTASLP20}, but now for the simpler VBx. 
This derivation should be much easier to follow for readers interested in this specific model. 
It also allows us to elaborate on how to make this simplified model much more computationally efficient.

All our code is made publicly available: the recipe for training the x-vector extractor (same architecture for both 8\,kHz and 16\,kHz), the trained extractors and the pipeline for applying BHMM diarization using agglomerative hierarchical clustering (AHC) as initialization~\url{https://github.com/BUTSpeechFIT/VBx}.


\section{VBx Diarization model}
\label{sec:themodel}

This section introduces the VBx model which is used in all the experiments in this paper.
The derivation of the inference formulae is also provided. 
While this derivation is essentially the same as the one that can be found in \cite{DiezTASLP20}, it only addresses the simple model used in this paper.
In fact, the following text is the same as Section II from  \cite{DiezTASLP20}, rewritten and simplified to address only the  model considered in this paper. 
We intentionally reused the text, structure and symbols from the mentioned paper for the following reasons: we want to make the treatment of the model here as self-contained as possible and we would like to facilitate the comparison of the simplified model with the more complex full model proposed in \cite{DiezTASLP20}.

\subsection{Model overview}

As described in the previous section, we expect a sequence of x-vectors extracted from consecutive short segments of speech
as input to our diarization method, which aims to cluster these x-vectors according to their speaker identity.

Our diarization model assumes that the input sequence of x-vectors is generated by an HMM with speaker-specific state distributions.
To facilitate the discrimination between speakers, the speaker- (or HMM state-) specific distributions are derived from a PLDA \cite{kenny10PLDA_HTP} model pre-trained on a large number of speaker-labeled x-vectors. 
More details on how the speaker-specific distributions are derived from the PLDA are given in section \ref{sec:speakerdist}. For now, it is sufficient to note that the speaker distributions will be represented only by a latent vector $\mathbf{y}_s$ of the same dimensionality as the x-vectors.

We use an ergodic HMM with one-to-one correspondence between the HMM states and the speakers, where transitions from any state to any state are possible. Note that our model does not consider any overlapped speech as each speech frame is assumed to be generated from an HMM state corresponding to only one of the $S$ speakers. The transition probabilities can be used to discourage too frequent transitions between speakers in order to reflect speaker turn durations of a natural conversation. More details on setting and learning the transition probabilities can be found in section~\ref{sec:topology}.
 
Let $\XX=\{\xx_1, \xx_2,...,\xx_T\}$ be the sequence of observed x-vectors and $\ZZ=\{z_1, z_2,...,z_T\}$ the corresponding sequence of discrete latent variables defining the hard alignment of  x-vectors to HMM states. In our notation, $z_t=s$ indicates that the speaker  (HMM state) $s$  is responsible for generating observation $\xx_t$.

To address the speaker diarization (SD) task using our model, the speaker distributions (i.e. the vectors $\mathbf{y}_s$) and the latent variables $z_t$ are jointly estimated given an input sequence $\XX$. The solution to the SD task is then given by the most likely sequence $\ZZ$, which encodes the alignment of speech frames to speakers.

\subsection{HMM topology}\label{sec:topology}

\begin{figure}[htb]
\centering
\begin{tikzpicture}
\tikzstyle{main}=[circle, minimum size = 6mm, thick, draw =black!80, node distance = 10mm]
\tikzstyle{small}=[circle, minimum size = 1mm, thick, draw =black!80, node distance = 8mm]
\tikzstyle{nonr}=[circle, thick, draw =black!80, node distance = 10mm]
\tikzstyle{connect}=[-latex, thick]
\tikzstyle{box}=[rectangle, draw=black!100]
  \node[main, fill = white!100,draw=black!80] (s1) [] {$\mathbf{s}_{1}$};
  \node[small] (s) [fill = black!100,below= of s1] {};
  \node[main] (s2) [below right=of s] {$\mathbf{s}_{2}$ };
  \node[main] (s3) [below left=of s] { $\mathbf{s}_3$};
  \node[nonr, fill = white!100,draw=none] (z1) [above right =of s1] {};
  \node[nonr, fill = white!100,draw=none] (z2) [above right =of s2] {};
  \node[nonr, fill = white!100,draw=none] (z3) [above left =of s3] {};
  \path (s1) edge [connect,  out=-120,in=135] node[left] {$1-P_\text{loop}$} (s)
		(s2) edge [connect,  out=100,in=-10] node[right=3pt] {$1-P_\text{loop}$} (s)
		(s3) edge [connect,  out=5,in=-100]  node[below  =13pt] {$1-P_\text{loop}$}(s)
        (s1) edge [loop, connect, out=70,in=110, looseness=5] node[ right=3pt]{$P_{loop}$} (s1)
        (s2) edge [loop, connect,  out=-70,in=-30, looseness=5] 
        node[right=3pt]{$P_{loop}$} (s2)
        (s3) edge [loop, connect, out=-160,in=-120, looseness=5] node[left=3pt]{$P_{loop}$} (s3)
        (s) edge [connect,  out=45,in=-60] node[right] {$\pi_1$} (s1)
		(s) edge [connect,  out=-80,in=170] node[right=1pt] {$\pi_2$} (s2)
		(s) edge [connect,  out=190,in=80]  node[left=3pt] {$\pi_3$}(s3);
\end{tikzpicture}
\caption{HMM model for 3 speakers with 1 state per speaker, with a dummy non-emitting (initial) state. }
\vspace{-2mm}
 \label{HMMmodel1state}
\end{figure}
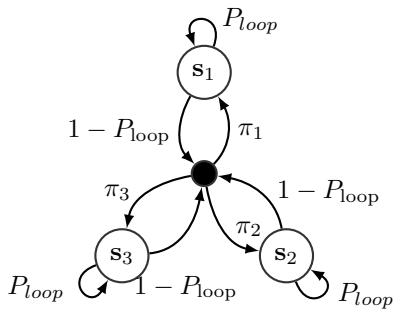

The HMM topology and transition probabilities model the speaker turn durations. The HMM model is ergodic (transitions between all states are possible).
Figure~\ref{HMMmodel1state} shows an example of the HMM topology for only $S=3$ speakers.  The transition probabilities are set as follows: we transition back to the same speaker/state with probability $P_{loop}$. 
This probability is one of the tunable parameters in the model.
The remaining probability $(1-P_{loop})$ is the probability of changing speaker, which corresponds to the transition to the non-emitting node in Figure \ref{HMMmodel1state}. 
From the non-emitting node, we immediately transition to one of the speaker states with probability $\pi_s$.\footnote{For convenience, we allow to re-enter the same speaker as it leads to simpler update formulae.} Therefore, the probability of leaving a speaker and entering another speaker $s$ is $(1-P_{loop})\pi_s$.
To summarize, the probability of transitioning from state $s'$  to state $s$ is
\begin{equation}
p(s|s')=(1-P_{loop})\pi_s+\delta(s=s') P_{loop}, 
\end{equation}
where $\delta(s=s')$ equals 1 if $s=s'$ and is $0$ otherwise.

The non-emitting node in Figure \ref{HMMmodel1state} is also the initial state of the model. Therefore, the probabilities $\pi_s$ also control the selection of the initial HMM state (i.e. the state generating the first observation).
These probabilities $\pi_s$ are inferred (jointly with the variables $\mathbf{y}_s$ and $z_t$) from the input conversation. Thanks to the automatic relevance determination principle~\cite{Bishop2006} stemming from our Bayesian model, zero probabilities will be learned for the $\pi_s$ corresponding to redundant speakers, which effectively drops such speakers from the HMM model. 
Typically, 
we initialize the HMM with a larger number of speakers (see section \ref{diarizationsetup}) and we make use of this behavior to drop the redundant speakers (i.e. to estimate the number of speakers in the conversation).
\subsection{Speaker-specific distributions}\label{sec:speakerdist}

The speaker (HMM state) specific distributions are derived from a PLDA which is a standard model used for comparing x-vectors in speaker verification \cite{kenny10PLDA_HTP}. 
Here, only a simplified variant of PLDA is considered, which is often referred to as \textit{two-covariance model} \cite{brummer2010spkpartitioning}.
This model assumes that the distribution of x-vectors specific to speaker $s$ is Gaussian $\mathcal{N}(\hat{\mathbf{x}}_t;\hat{\mathbf{m}}_s,\boldsymbol{\Sigma}_{w})$, where $\boldsymbol{\Sigma}_{w}$ is the within-speaker covariance matrix shared by all speaker models, and $\hat{\mathbf{m}}_s$ is the speaker-specific mean.
Speaker means are further assumed to be Gaussian distributed $\mathcal{N}(\hat{\mathbf{m}}_s; \mathbf{m}, \boldsymbol{\Sigma}_{b})$, where $\mathbf{m}$ is the global mean and $\boldsymbol{\Sigma}_{b}$ is the between-speaker covariance matrix.
In general, $\boldsymbol{\Sigma}_{w}$ and $\boldsymbol{\Sigma}_{b}$ can be full covariance matrices. However, to further simplify and speed up the inference in our model, we assume that the x-vectors are linearly transformed into a space where $\boldsymbol{\Sigma}_{b}$ is diagonal and $\boldsymbol{\Sigma}_{w}$ is identity. This can be achieved as follows:

Let $\hat{\XX}$ be the matrix of original (untransformed) x-vectors that the parameters of the original PLDA model $\mathbf{m},$ $\boldsymbol{\Sigma}_{w}$ and $\boldsymbol{\Sigma}_{b}$  were estimated on.
The x-vectors that are used as input for the diarization algorithm are obtained as 
\begin{equation}
\label{eq:transformxvec}
    \XX=(\hat{\XX}-\mathbf{m})\mathbf{E}
\end{equation} 
\noindent where $\mathbf{E}$ is the transformation matrix which transforms the x-vectors into the desired space.
This matrix can be obtained by solving the standard generalized eigen-value problem
\begin{equation}
    \boldsymbol{\Sigma}_{b} \mathbf{E} = \boldsymbol{\Sigma}_{w} \mathbf{E} \boldsymbol{\Phi} 
\end{equation}
\noindent where $\mathbf{E}$ is the matrix of eigen-vectors and $\boldsymbol{\Phi}$ is the diagonal matrix of eigen-values, which is also the between-speaker covariance matrix in the transformed space.
Note that the eigen-vectors $\mathbf{E}$ are, in fact, bases of linear discriminant analysis (LDA) estimated directly from the PLDA model parameters. Therefore, if we construct $\boldsymbol{\Phi}$ only using $R$ largest eigen-values and assemble $\mathbf{E}$ only using the corresponding eigen-vectors, \eqref{eq:transformxvec} further performs LDA dimensionality reduction of x-vectors to $R$-dimensional space. We use $R$ as one of the hyper-parameters of the VBx method. In equation~\eqref{eq:transformxvec}, we have also subtracted the global mean from the original x-vectors to have the new set of x-vectors  zero-centered.

In summary, the PLDA model compatible with the new set of x-vectors $\XX$ suggests that speaker-specific means are distributed as 
\begin{equation}
p(\mathbf{m}_s) = \mathcal{N}(\mathbf{m}_s; \mathbf{0}, \boldsymbol{\Phi}).
\end{equation}

For convenience and for the compatibility with the notation introduced in \cite{DiezTASLP20}, we further re-parametrize the speaker mean as 
\begin{equation}
\mathbf{m}_s=\mathbf{V} \mathbf{y}_s,
\label{eq:PLDA}
\end{equation}
\noindent where diagonal matrix $\VV=\boldsymbol{\Phi}^{\frac{1}{2}}$  and $\mathbf{y}_s$ is a standard normal distributed random variable
\begin{equation}
\label{eq:normalys}
    p(\yy_s)=\mathcal{N}(\yy_s;\mathbf{0},\mathbf{I}).
\end{equation}
The speaker-specific distribution of x-vectors is
\begin{equation}
p(\mathbf{x}_t|\mathbf{y}_s) = \mathcal{N}(\mathbf{x}_t ; \mathbf{V} \mathbf{y}_s, \mathbf{I}),
\label{eq:datcondspkSIMP}
\end{equation}
\noindent where $\mathbf{I}$ is identity matrix.

In our diarization model, we use \eqref{eq:datcondspkSIMP} to model the speaker (HMM state) distributions. 
This distribution is fully defined only in terms of the speaker vector $\yy_s$ (and the pre-trained matrix $\VV$ shared by all the speakers).
The speaker vector $\yy_s$ is treated as a latent variable with standard normal prior~\eqref{eq:normalys}, which is why the BHMM model is called \textit{Bayesian}\footnote{However, unlike other ``Fully Bayesian'' HMM implementations~\cite{Fox09Sticky,BealVBLowerBThesis}, we do not impose any prior on the transition probabilities.}.
This way, the full PLDA model is incorporated into the BHMM in order to properly model between- and across-speaker variability. Therefore, the model is capable of discriminating between speakers just like PLDA model when used for speaker verification. 

\subsection{Bayesian HMM}\label{sec:BayesianHMM}

To summarize, our complete  model for SD is a Bayesian HMM, which is defined in terms of the state-specific distributions (or so-called output probabilities)
\begin{equation}
p(\mathbf{x}_t|z_t=s) =  p(\mathbf{x}_t|s) =  p(\mathbf{x}_t|\mathbf{y}_s)
 \end{equation}
described  in section~\ref{sec:speakerdist}
and the transition probabilities
\begin{equation}
p(z_t=s|z_{t-1}=s')  = p(s|s')
 \end{equation}
described in section~\ref{sec:topology}.
By abuse of notation, $p(z_1|z_0)$ will correspond to the initial state probability $p(z_1{=}s)=\pi_s$ in the following formulae.

The complete model can be also defined in terms of the joint probability of the observed and latent random variables (and their factorization) as
\begin{flalign}
\label{jointprobeq}
 p  (\XX,\ZZ,\YY)&= p(\XX|\ZZ,\YY) p(\ZZ) p(\YY)\\
&\nonumber =
\prod_t p\left(\xx_t|z_t\right)
\prod_t p\left(z_t|z_{t-1}\right)
\prod_s  p\left(\yy_s\right),
\end{flalign}
where $\YY=\{\yy_1, \yy_2,...,\yy_S\}$ is the set of all the speaker-specific latent variables.

The model assumes that each x-vector sequence corresponding to an input conversation is obtained using the following generative process:
\begin{mdframed}[
    align=center,
    linecolor=black,
    linewidth=0.6pt,
    userdefinedwidth=0.7\columnwidth,
]
\begin{algorithmic}
  \FOR{$s=1..S$}
      \STATE $\mathbf{y}_s \sim \mathcal{N}(0,\mathbf{I})$
  \ENDFOR
  \FOR{$t=1..T$}
  \STATE $z_t \sim p(z_t|z_{t-1})$
  \STATE $\mathbf{x}_t \sim p(\mathbf{x}_t|z_t)$
  \ENDFOR
\end{algorithmic}
\end{mdframed}

\subsection{Diarization inference}

\label{sec:inference}
The diarization problem consists in finding the assignment of frames to speakers, which is represented by the latent sequence $\ZZ$. In order to find the most likely sequence $\ZZ$, we need to infer the posterior distribution  $p(\ZZ|\mathbf{X})=\int p(\ZZ,\mathbf{Y}|\mathbf{X}) d\mathbf{Y}$. 
Unfortunately, the evaluation of this integral is intractable, and therefore, we will approximate it using variational Bayes (VB) inference \cite{Bishop2006}, where the distribution $p(\ZZ,\YY|\mathbf{X})$ is approximated by $q(\ZZ,\YY)$. We use the mean-field approximation \cite{Bishop2006,KennyBayDiar} assuming that the approximate posterior distribution factorizes as
\begin{equation}
q(\ZZ,\YY) = q(\ZZ)q(\YY).
\label{eq:meanfieldfactorization}
\end{equation}
The particular form of the approximate distributions $q(\ZZ)$ and $q(\YY)$ directly follows from the optimization described below.

We search for such $q(\ZZ,\YY)$ that minimizes the Kullback-Leibler divergence $D_{KL}(q(\ZZ,\YY)\|p(\mathbf{Z},\mathbf{Y}|\mathbf{X}))$, which is equivalent to maximizing the standard VB objective -- the evidence lower bound objective (ELBO) \cite{Bishop2006}
\begin{equation}\label{ELBO}
\begin{split}
\mathcal{L}\left(q(\mathbf{\XX,\YY})\right)=&E_{q(\YY,\ZZ)}\left\lbrace \ln\left(\frac{p(\XX,\YY,\ZZ)}{q(\YY,\ZZ)}\right)\right\rbrace.
\end{split}
\end{equation}
Using the factorization~\eqref{eq:meanfieldfactorization}, the ELBO can be split into three terms
\begin{equation}\label{fetchELBO}
\begin{split}
\hat{\mathcal{L}}&\left(q(\XX,\YY)\right)= F_A E_{q(\YY,\ZZ)}\left[ \ln p(\XX|\YY,\ZZ)\right]\\
& \qquad\quad +F_B E_{q(\YY)}\left[ \ln\frac{p(\YY)}{q(\YY)}\right] + E_{q(\ZZ)}\left[ \ln\frac{p(\ZZ)}{q(\ZZ)}\right], 
\end{split}
\end{equation}
where the first term is the expected log-likelihood of the observed x-vector sequence $\XX$ and the second and third terms are Kullback-Leibler divergences $D_{KL}(q(\YY)\|p(\YY))$ and $D_{KL}(q(\ZZ)\|p(\ZZ))$ regularizing the approximate posterior distributions $q(\YY)$ and $q(\ZZ)$ towards the priors $p(\YY)$ and $p(\ZZ)$. In~\eqref{fetchELBO}, we modified the ELBO by scaling the first two terms by constant factors $F_A$ and $F_B$.\footnote{Note that similar scaling factor for the third term would be redundant as only the relative scale of the three factors is relevant for the optimization.} The theoretically correct values for these factors leading to the original ELBO~\eqref{fetchELBO} are $F_A=F_B=1$. However, choosing different values gives us finer control over the inference, which can be used to improve diarization performance. For further details on the specific effects these scaling factors have in the inference, we refer the reader to \cite{DiezTASLP20}.

As described above, we search for the approximate posterior $q(\ZZ,\YY)$ that maximizes the ELBO~\eqref{fetchELBO}. In the case of the mean-field factorization~\eqref{eq:meanfieldfactorization}, we proceed iteratively by finding the $q(\YY)$ that maximizes the ELBO given fixed $q(\ZZ)$ and vice versa. This section provides all the formulae necessary for implementing these updates or for understanding our open-source Python implementation\footnote{\url{http://speech.fit.vutbr.cz/software/vb-diarization-eigenvoice-and-hmm-priors}}. 
In this section, we do not give any details on deriving the update formulae. For the readers interested in the derivations, we prepared a technical report \cite{DiezVBxreport21}.

\subsubsection{Updating $q(\YY)$}

Given a fixed $q(\ZZ)$, the distribution over $\YY$ that maximizes the ELBO is 
\begin{equation}
q^*(\YY) = \prod_s q^*(\yy_s),
\end{equation}
where the speaker-specific approximate posteriors
\begin{equation}
q^*(\yy_s)= \mathcal{N}\left(\mathbf{y}_s|\boldsymbol{\alpha}_s,\mathbf{L}_s^{-1}\right)
\label{YVBupdate}
\end{equation}
are Gaussians with the mean vector and precision matrix
\begin{align}
\boldsymbol{\alpha}_s=\frac{F_A}{F_B}\mathbf{L}_s^{-1}\sum_t \gamma_{ts}\boldsymbol{\rho}_t\label{YVBupdate2_alpha}\\ \mathbf{L}_s=\mathbf{I}+\frac{F_A}{F_B}\left(\sum_t \gamma_{ts}\right)\boldsymbol{\Phi}.
\label{YVBupdate2_L}
\end{align}
\noindent where 
\begin{equation}
\label{eq:rho}
    \boldsymbol{\rho}_t= \VV^T \xx_t
\end{equation}
In this update formula, $\gamma_{ts}=q(z_{t}=s)$ is the marginal approximate posterior derived from the current estimate of the distribution $q(\ZZ)$ (see below), which can be interpreted as the responsibility of speaker $s$ for generating observation $\xx_t$ (i.e. defines a soft alignment of x-vectors to speakers). 

If we compare these update formulae to the corresponding ones from the more complex BHMM model in \cite{DiezTASLP20}, it can be seen that in \cite{DiezTASLP20}, $\boldsymbol{\Phi}_t$ is a frame-dependent full-matrix computationally expensive to calculate. In contrast, $\boldsymbol{\Phi}$ here does not depend on time frame $t$ and, as pointed out in section \ref{sec:speakerdist}, it is a diagonal matrix.
Therefore also matrix $\mathbf{L}_s$ is diagonal, and its inversions and application in \eqref{YVBupdate2_alpha} become trivial.

\subsubsection{Updating $q(\ZZ)$}
We never need to infer the complete distribution over all the possible alignments of observations to speaker $q(\ZZ)$. When updating $q(\YY)$ using~\eqref{YVBupdate2_alpha} and~\eqref{YVBupdate2_L}, we only need the marginals $\gamma_{ts}=q(z_{t}=s)$. Therefore, when updating $q(\ZZ)$, we can directly search for the responsibilities $\gamma_{ts}$ that correspond to the distribution $q^*(\ZZ)$ maximizing the ELBO given a fixed $q(\YY)$. Similar to the standard HMM training, such responsibilities can be calculated efficiently using a forward-backward algorithm as
\begin{equation}
\gamma_{ts}=\frac{A(t,s)B(t,s)}{\overline{p}(\XX)}
\label{spkrespons}
\end{equation}
where the forward probability
\begin{equation}
\begin{split}
\label{forward}
A(t,s)=\bar{p}(\xx_t|s)\sum_{s'}A(t-1,s')p(s|s')
\end{split}
\end{equation}
is recursively evaluated by progressing forward in time for ${t{=}1..T}$ starting with $A(0,s)=\pi_s$. Similarly,
\begin{equation}
\begin{split}
\label{backward}
B(t,s)=\sum_{s'}B(t+1,s')\bar{p}(\xx_{t+1}|s')p(s'|s)
\end{split}
\end{equation}
is the backward probability evaluated using backward recursion for times $t=T..1$ starting with $B(T,s)=1$. 
\begin{equation}
\label{totalA}
\overline{p}(\XX)=\sum_s A(T,s)
\end{equation}
is the total forward probability and
\begin{align}
\label{almostHMMp}
\nonumber
\log \overline{p}(\xx_t|s) = & F_A \left[  \boldsymbol{\alpha}_s^T\boldsymbol{\rho}_t-\frac{1}{2}\tr\left(\boldsymbol{\Phi}\left[\mathbf{L}_s^{-1}+\boldsymbol{\alpha}_s\boldsymbol{\alpha}_s^T\right]\right) -\frac{D}{2}\ln 2\pi -\frac{1}{2}\xx_t^T\xx_t \right] \\
= & F_A \left[  \boldsymbol{\alpha}_s^T\boldsymbol{\rho}_t-\frac{1}{2}\boldsymbol{\phi}^T\left[\boldsymbol{\lambda}_s+\boldsymbol{\alpha}_s^2\right]  -\frac{D}{2}\ln 2\pi -\frac{1}{2}\xx_t^T\xx_t \right]
\end{align}

\noindent is the expected log likelihood of observation $\xx_t$ given a speaker $s$ taking into account its uncertainty $q(\yy_s)$. The second line of \eqref{almostHMMp} corresponds to an efficient evaluation of this term, where vector $\boldsymbol{\phi}$ is the diagonal of the diagonal matrix $\boldsymbol{\Phi}$, vector $\boldsymbol{\lambda}$ is the diagonal of the diagonal matrix $\mathbf{L}_s^{-1}$ and the square in $\boldsymbol{\alpha}_s^2$ is element-wise. Note also that the terms $-\frac{D}{2}\ln 2\pi -\frac{1}{2}\xx_t^T\xx_t$ in \eqref{almostHMMp} are not only constant over VB iterations, but also constant for different speakers $s$. As a consequence, contribution of these terms cancels in \eqref{spkrespons} and therefore does not have to be calculated at all.

\subsubsection{Updating $\pi_s$} Finally, the speaker priors $\pi_s$ are updated as maximum likelihood type II estimates \cite{Bishop2006}: Given fixed $q(\YY)$ and $q(\ZZ)$, we search for the values of $\pi_s$ that maximize the ELBO \eqref{fetchELBO}, which gives the following update formula
\begin{equation}
\pi_s \propto\ \gamma_{1s}+ \frac{(1{-}P_{loop})\pi_s}{\overline{p}(\XX)}\sum_{t=2}^T \sum_{s'} A(t{-}1,s') p(\xx_t|s) B(t,s)
\label{thepiupdate}
\end{equation}
with the constraint $\sum_s \pi_s=1$.
As described in section \ref{sec:topology}, this update tends to drive the $\pi_s$ corresponding to ``redundant speakers'' to zero values, which effectively drops them from the model and selects the right number of speakers in the input conversation.

\subsubsection{Evaluating the ELBO}
The convergence of the iterative VB inference can be monitored by evaluating  the ELBO objective. For the Bayesian HMM, the ELBO can be efficiently evaluated (see page 95 of \cite{BealVBLowerBThesis}) as
\begin{equation}
\hat{\mathcal{L}} = \ln\overline{p}(\XX) +\sum_s \frac{F_B}{2} \left(R+\ln|\mathbf{L}_s^{-1}|-\tr(\mathbf{L}_s^{-1})-\boldsymbol{\alpha}_s^T\boldsymbol{\alpha}_s\right),
\label{lowerbound}
\end{equation}
where $R$ is the dimensionality of the x-vectors. Note, that since $\mathbf{L}_s$ is a diagonal matrix, $\ln|\mathbf{L}_s^{-1}|$ can be calculated just as the sum of the log of the elements in the diagonal.
This way of evaluating the ELBO is very practical as the term $\overline{p}(\XX)$ from~\eqref{totalA} is obtained as a byproduct of ``updating $q(\ZZ)$'' using the forward-backward algorithm. On the other hand,~\eqref{lowerbound} allows to evaluate the ELBO only right after the $q(\ZZ)$ update. It does not allow to monitor the improvement in ELBO obtained form $q(\YY)$ or $\pi_s$ updates, which might be useful for debugging purposes. Therefore, we also provide the derivation formulae for the explicit evaluation of all three ELBO terms from~\eqref{fetchELBO} in \cite{DiezVBxreport21}.

The complete VB inference consisting of iterative updates of $q(\YY)$, $q(\ZZ)$ and parameters $\pi_s$ is summarized in the following algorithm:
\begin{mdframed}[
    align=center,
    linecolor=black,
    linewidth=0.6pt,
    userdefinedwidth=0.9\columnwidth,
]\begin{algorithmic}
  \STATE Initialize all $\gamma_{ts}$ as described in section~\ref{diarizationsetup}.
  \REPEAT  
    \STATE   Update $q(\mathbf{y}_s)$ for $s{=}1..S$ using \eqref{YVBupdate}
    \FOR{$t=1..T$}
    \STATE Calculate $A(t,s)$ for $s{=}1..S$ using \eqref{forward}
    \ENDFOR
    \FOR{$t=T..1$}
    \STATE Calculate $B(t,s)$ for $s{=}1..S$ using  \eqref{backward}
    \ENDFOR
    \STATE Update $\gamma_{ts}$ for $t{=}1..T, s{=}1..S$ using \eqref{spkrespons}
    \STATE Update $\pi_s$ for $s{=}1..S$ using \eqref{thepiupdate}
    \STATE Evaluate ELBO $\hat{\mathcal{L}}$ using \eqref{lowerbound}
    \UNTIL convergence of $\hat{\mathcal{L}}$
\end{algorithmic}
\end{mdframed}


\section{Experimental setup}

\subsection{x-vector extractor and PLDA}
\label{sec:xvector}

As described in the previous section, VBx diarization relies on a pre-trained x-vector extractor and a PLDA model. Since we report results on both 16\,kHz recordings (DIHARD and AMI) and 8\,kHz telephone recodings (CALLHOME), we train two x-vector extractors and the corresponding PLDA models, one for each condition. The complete PyTorch~\cite{paszke2019pytorch} recipe for x-vector extractor and PLDA training is available at \url{https://github.com/phonexiaresearch/VBx-training-recipe}.

\subsubsection{x-vector extractor architecture}
Both 8\,kHz and 16\,kHz x-vector extractors use the same deep neural network architecture based on ResNet101~\cite{he2016deep, zeinali2019but}.
In both cases, the neural network inputs are 64 log Mel filter bank features extracted every 10\,ms using 25\,ms window. The two x-vector extractors differ only in the frequency ranges spanned by the Mel filters, which are 20-7700\,Hz and 20-3700\,Hz for the 16\,kHz and 8\,kHz systems, respectively. The x-vector extractor architecture is summarized in Table~\ref{tab:resnet101}. The first 2D convolutional layer operates on the $64\times T$ matrix of log Mel filter bank features, where $T$ is the number of frames in the input segments. For training, we use 4\,s segments (i.e. $T=400$).  The following layers are standard ResNet blocks \cite{he2016deep}. As in the original x-vector architecture~\cite{snyder2018x}, the statistical pooling layer is used to aggregate information over the whole speech segment (i.e. mean and standard deviation of activations is calculated over the time dimension).
After the pooling layer, a linear transformation is used to reduce the dimensionality to obtain the $256$-dimensional x-vectors.
\begin{table}[!th]
\renewcommand{\arraystretch}{0.85}
\caption{\label{tab:resnet101} \textit{The structure of the proposed ResNet101 architecture. The first dimension of the input shows the size of the filterbank and the second dimension indicates the number of frames.}}
  \centerline{
  \setlength\tabcolsep{4pt}
    \begin{tabular}{l l l l}
    \toprule
    \textbf{Layer} & \textbf{Structure} & \textbf{Stride} & \textbf{Output} \\
    \midrule
    Input & - & - & $64 \times $T$ \times 1$ \\
    Conv2D-1 & $3 \times 3, 32$ & 1 & $64 \times $T$ \times 32$ \\
    \midrule
    ResNetBlock-1 & $\myatopthree{1 \times 1, 32}{3 \times 3, 32}{1 \times 1, 128} \times 3$ & 1 & $64 \times $T$ \times 128$ \\
    ResNetBlock-2 & $\myatopthree{1 \times 1, 64}{3 \times 3, 64}{1 \times 1, 256} \times 4$ & 2 & $32 \times $T/2$ \times 256$ \\
    ResNetBlock-3 & $\myatopthree{1 \times 1, 128}{3 \times 3, 128}{1 \times 1, 512} \times 23$ & 2 & $16 \times $T/4$ \times 512$ \\
    ResNetBlock-4 & $\myatopthree{1 \times 1, 256}{3 \times 3, 256}{1 \times 1, 1024} \times 3$ & 2 & $8 \times $T/8$ \times 1024$ \\
    \midrule
    Statistics Pooling & - & - & $16 \times 1024$ \\
    Flatten & - & - & $16384$ \\
    Linear & - & - & $256$ \\
   \bottomrule 
   \end{tabular}
}       
\end{table}

The x-vector extractors are trained using stochastic gradient descent and additive angular margin loss~\cite{deng2019arcface} with speaker identities as class labels. We ramp-up the margin during the first two epochs (pass through the training data) and then train the neural network for another epoch with fixed margin $m=0.2$.
 
 \subsubsection{x-vector extractor training data}
The 16\,kHz x-vector extractor is trained using data from VoxCeleb1~\cite{Nagrani17} (323\,h of speech from 1211 speakers), VoxCeleb2~\cite{Chung18b} (2290\,h, 5994 speakers) and CN-CELEB~\cite{fan2020cn} (264\,h, 973 speakers). The energy-based VAD from Kaldi~\cite{povey2011kaldi} toolkit is used to remove silence frames. Speakers with less than 2 recordings are discarded. Further, we drop utterances with less than 4 seconds of speech. This way, about 4\% of speech data is discarded. Data augmentation is performed the same way as in the SRE16 Kaldi recipe \cite{snyder_kaldi_recipe}. This way, we obtain four additional copies of the data with artificially added noise, music or reverberation. 
Training examples are randomly sampled from the training data. This way we extract about 89 million examples (original and augmented 4s segments), which cover more than 60\% of the speech from the training corpora.

To train the 8\,kHz x-vector extractor, the same data sets are used as in the 16\,kHz case. Additionally, the following data sets were used: 
Mixer collection (NIST SRE 2004-2010, 3805\,h, 4254 speakers), Switchboard (1170\,h, 2591 speakers) and DeepMine~\cite{zeinali2018deepmine} (688\,h, 1858 speakers). Any wide-band data used were downsampled to 8\,kHz and passed through a telephone codec. The same data selection and augmentation was used as for the 16\,kHz case.
Note that about 30\% of DeepMine data were discarded as this dataset contains many utterances with less than 4 seconds of speech (mostly phrases for text-dependent speaker verification). 

\subsubsection{PLDA training}
The 8\,kHz and 16\,kHz PLDA models are trained on the same data as the corresponding x-vector extractors. For this purpose, one x-vector is extracted from each individual recording (e.g. one cut from a YouTube video in the case of the VoxCeleb data). The length of such recordings can range from 4\,s to several minutes. Note that the PLDA trained on such x-vectors is later used in VBx to operate on x-vectors extracted from much shorter 1.5\,s segments. This mismatch, however, does not seem to negatively affect diarization performance.

\subsection{Diarization pipeline}
\label{diarizationsetup}
To perform the diarization, each input recording is first split into speech segments according to the oracle VAD and the segments shorter than 0.1\,s are discarded. From these segments, x-vectors are extracted every 0.25\,s from overlapping sub-segments of 1.5\,s (or less than 1.5\,s for the last sub-segments or shorter segments). The x-vectors are centered, whitened and length normalized \cite{Romero11lengthnorm} (which is also done for the PLDA training data).

As described in section~\ref{sec:topology}, VBx diarization needs an initial assignment of x-vectors to speakers.
For this purpose, the x-vectors are pre-clustered using AHC to obtain the initial speaker labels. The only input to the AHC is the matrix of cosine similarities between all pairs of x-vectors. The threshold used as the stopping criterion for AHC is tuned to under-cluster so that the following VBx has more freedom to search for the optimal results and converge to the right number of speaker models\footnote{Note that the inference in BHMM cannot converge to higher number of speakers than what is suggested by the AHC-based initialization.}. Nevertheless, the same threshold is used for all our results on all the datasets.

In the final step, the x-vectors are further clustered using the VBx model and the inference described in section~\ref{sec:themodel}. For this step, the x-vector dimensionality is further reduced to 128 dimensions (see parameter $R$ in section~\ref{sec:speakerdist}). Note that unlike in our previous works~\cite{landini2020but,diez2020optimizing}, we do not perform any adaptation of PLDA models to the target data as this paper aims to present a simple diarization system which performs well for different datasets. Nevertheless, we tune the VBx parameters: $F_A$, $F_B$, $P_{loop}$ on the respective DIHARDII, AMI and CALLHOME development sets.

In order to demonstrate the effectiveness of the VBx method, we also report results for baseline systems where only AHC is used to cluster x-vectors. In this case, the stopping threshold is tuned to obtain the best performance on the respective development set.

\subsection{Evaluation Datasets}

\subsubsection{CALLHOME}

The 2000 NIST Speaker Recognition Evaluation (LDC2001S97\footnote{\url{https://catalog.ldc.upenn.edu/LDC2001S97}}) dataset, usually referred as ``CALLHOME'', \cite{NISTSRE2000evalplan} has been the standard dataset for diarization in the last decade \cite{Shum13,Senoussaoui2014MS,Garcia-RomeroSS17}. In its full form\footnote{Not only the English partition, nor the partition limited to 2 speaker audios sometimes used.}, it consists of 499\footnote{One audio is commonly excluded because its references have formatting errors} recordings of conversational telephone speech in Arabic, English, German, Japanese, Mandarin and Spanish. The number of speakers per recording ranges between 2 and 7, although 87\% of the files contain only 2 or 3 speakers. It amounts to around 15 hours of speech after VAD.

Since a development-evaluation split for CALLHOME is not available, we split the dataset into two halves as defined in the Kaldi recipe for CALLHOME\footnote{\url{https://github.com/kaldi-asr/kaldi/blob/master/egs/callhome\_diarization/v2/run.sh}}. We use this split to perform cross-validation to tune parameters i.e. $F_A$, $F_B$ and $P_{loop}$. 

\subsubsection{AMI corpus}

\noindent When trying to cover the most standard datasets for speaker diarization, we could not leave AMI out
\cite{carletta2005ami}.
The AMI meeting corpus is a multi-modal data collection of 100 hours of meeting recordings.
This corpus was recorded using both close-talking and far-field microphones. It consists of 171 meetings recorded at the University of Edinburgh (U.K.), Idiap (Switzerland), and the TNO Human Factors Research Institute (The Netherlands).
The dataset comes with annotations for automatic speech recognition (ASR).
AMI has been widely used by the community for diarization purposes. 
Still, somewhat surprisingly, authors do not use a standard evaluation protocol to report the results on this dataset.

The description of the different evaluation protocols reported in the literature as well as our proposed new protocol for evaluation on AMI can be found in section \ref{sec:Amiprotocols}.

\subsubsection{DIHARD II}

DIHARD II is one of the newest datasets designed for diarization. This dataset was created as an extension of the first DIHARD dataset for the second DIHARD challenge \cite{DIHARD19}, the second of a yearly series of challenges designed to foster research on diarization in hard conditions. One of the main features of this dataset, is that it contains audios from several sources (YouTube, court rooms, meetings, etc.) covering a wide range of numbers of speakers per recording (1 to 10) and large variety of channels and audio conditions.
The corpus consists of 192 development and 194 evaluation recordings, containing around 18\,h and 17\,h of speech, respectively.

\subsection{Evaluation protocol}
\label{sec:eval}
Diarization performance is evaluated in terms of diarization error rate (DER) as defined by NIST \cite{NISTRT}:
\begin{equation}
    DER=\frac{SER+FA+Miss}{Total\_speech}
\end{equation}
\noindent where:
\begin{itemize}
    \item $SER$ stands for speaker error, the amount of time that speech is attributed to incorrect speakers
    \item $FA$ is false alarm, the amount of time that  non-speech regions are incorrectly attributed to a speaker (or time when overlapped speech is found in single speaker speech regions)
    \item $Miss$ stands for missed speech, the amount of time that speech is not attributed to any speaker
    \item Total\_speech is the total amount of speech, accounting also for speaker overlaps.
\end{itemize}
Note that, as we use oracle VAD in all our experiments, $FA$ error is zero, and the $Miss$ (due only to non-handled overlapped speech regions) can be directly calculated as $DER-SER$ as we also report SER. 
We also evaluate the system in terms of Jaccard error rate (JER), which has been established as secondary metric in the latest diarization challenges \cite{DIHARD19,chung2020spot}. JER is similar to DER, although it weighs every speaker equally, regardless of the amount of speech they produced.
All experiments are evaluated using the dscore tool\footnote{\url{https://github.com/nryant/dscore}}.

For a more thorough analysis of results, for CALLHOME and AMI we consider three setups for evaluation: 
First, a \textit{forgiving} one in which a 0.25\,s collar is considered for DER estimation and no overlap is evaluated. This is the standard configuration used for these datasets and allows a comparison of results with previously published works.
Second, we consider a similar evaluation using a 0.25\,s collar but accounting for overlapped speech, as in \cite{chung2020spot}. This \textit{fair}  setup covers a pragmatic scenario where all speech is evaluated, while being flexible on the speaker change points, as no realistic human annotation can achieve frame-precision. 
Finally, the \textit{full} one, in which no collar is used and overlapped speech is evaluated, which is in line with the setup used in latest diarization challenges \cite{DIHARD19,watanabe2020chime6}.
We consider that looking into the numbers for the \textit{fair} and \textit{full} setups is truly relevant on these datasets: very low DER values are already being achieved on these sets with the \textit{forgiving} setup, which suggest that future evaluations could shift into more challenging setups. Also, the hot-topic end-to-end approaches are likely to surpass the performance of current systems on  overlapped speech regions, and suitable baselines need to be established.

For the more recent DIHARD II dataset, only the  \textit{fair} and \textit{full} evaluation setups are considered.

Note that JER considers no collar and evaluates overlap regions by definition, so it is not affected by these configurations. To avoid confusion (and repetition), we only report this value on the \textit{full} setup.

\section{AMI evaluation protocols}
\label{sec:Amiprotocols}

\subsection{Evaluation protocols found in previously published works}
\label{sec:amiliterature}
As pointed out before, when it comes to evaluation on AMI, there is not a defined evaluation protocol.
Different authors evaluate their systems on different audio types (recordings from Mix-Headset microphones, microphone arrays, etc.). 
Besides, authors use different data partitions (train/dev/eval sets) and 
use different references.
This makes it practically impossible to compare results between sites. We would like to highlight that during our search for baselines in the literature, we found that most works are unaware of this inconsistency of evaluation protocols for AMI, which frequently leads to unfair comparisons.

Based on our literature review we replicated some evaluation protocols from works presenting remarkable performance. We use these protocols to evaluate our approach and fairly compare it with the respective works.

There are two major publicly available recipes for diarization on AMI that are included in Pyannote\footnote{\url{https://github.com/pyannote/pyannote-audio/tree/master/tutorials/pipelines/speaker_diarization}} \cite{Bredin2020} and Kaldi\footnote{\url{https://github.com/kaldi-asr/kaldi/tree/master/egs/ami/s5c}} \cite{povey2011kaldi} toolkits. 
These two recipes evaluate only on one audio type, which is the independent headset microphone mixed audio (Mix-Headset in AMI). 
Each recipe also derives their references in different ways from the official ASR transcriptions.
Finally, they both use their own partitioning of the data, Pyannote uses the Full-corpus AMI partition and Kaldi claims to use the official Full-corpus-ASR AMI partition\footnote{\label{footAMIpartitions}\label{amipartitions}\url{http://groups.inf.ed.ac.uk/ami/corpus/datasets.shtml}}. Nevertheless, Kaldi partition differs from the Full-corpus-ASR one, as it includes one meeting in the training set (IB4005) which causes speaker overlap between train and dev sets. This meeting is explicitly excluded in the Full-corpus-ASR partition. 
 
All the works that we compare with use different combinations of partitions and references taken mostly from these two recipes. 
The first six columns in Table~\ref{crazyAMI} summarize this mixture of protocols. Some works use Pyannote partition \cite{Bredin2020,bullock2019overlapaware}, while others use the Kaldi one \cite{sun2020combinationCambJour,Maciejewski2018AMI,raj2020multiclass,raj2020doverlap}, or even a modified version of the Kaldi one, excluding TNO meetings \cite{pal2020metalearning,Sun19CambridgeKaldiTNO}. 
As for references, AMI corpus comes only with manual and automatic transcriptions for ASR training and there are different ways of deriving diarization references out of them, as will be described later. In previously published works, we find again references from Pyannote \cite{Bredin2020,bullock2019overlapaware,raj2020multiclass,raj2020doverlap} and the ones from Kaldi \cite{Maciejewski2018AMI}, although some works use modified versions of these based on their own ASR forced alignment \cite{sun2020combinationCambJour}, or simply derive their own  \cite{pal2020metalearning}.
Note also the mix of evaluation setups (better described in section \ref{sec:eval}) with different collars and criteria for including or excluding overlapped speech. 
Column 5 in the table shows the amount of speech (in seconds) evaluated with each protocol for dev and eval sets. As it can be seen, evaluation can range from around 13000 to up to 52000 seconds of speech for the eval set depending on the evaluation setup used.

We strongly believe a standard evaluation protocol should be established on AMI.
Next, we introduce what we believe should be established as this new standard.

\subsection{New AMI evaluation protocol}
\label{newAMIprotocol}
The following evaluation protocol was built after discussions with researchers from different labs, authors of official Kaldi and Pyannote recipes (BUT, CLSP JHU, IRIT).

We propose to use the official Full-corpus-ASR partition\textsuperscript{\ref{footAMIpartitions}}. This way, we make the scoring of diarization tasks consistent with the scoring of speaker-attributed ASR.
As mentioned before, such partition is very similar to the Kaldi partition (in fact, dev and eval sets are the same) but it has no speaker overlap between sets, which makes it suitable for diarization tasks.

Regarding the references, authors of other works do not make clear how their references were created.
Our diarization references are directly derived from the AMI manual annotations, version 1.6.2.\footnote{\url{http://groups.inf.ed.ac.uk/ami/download/}}
These annotations are human transcriptions of all the meetings, containing words, vocal sounds and punctuation marks. To generate the references:
\begin{itemize}
    \item All words are considered as speech and included in the references. 
    \item Sounds of very different nature were annotated as vocal sounds. Some examples are Dutch speech, whistling, yawn, laughter, cough, clicking with tongue, raspberry noise, blowing nose, clapping, etc. 
    Some of these are clearly speech (Dutch speech), while others are clearly noises (blowing nose, clapping, etc.).
    For sounds such as laughter or whistling it is simply unclear if they should be considered for diarization purposes, as this would depend on the particular application of the system.
    Besides, we found that several of these vocal sounds are labeled without time annotations, which makes it impossible to add them to the references.
    We therefore decided to take a well defined, consistent and conservative approach in which all vocal sounds are discarded. This way, only the words that could be recognized by an ASR system are considered in our references.
    This is also more consistent with the task of speaker-attributed ASR.
    \item Speaker turns respect precisely the annotations,
    but adjacent speech segments (words) of the same speaker are merged not to create false ``break'' points.
    Consider the following example of an ASR transcription with speech from a single speaker which, as the AMI one, is composed of short segments of speech for each word: 

\begin{tcolorbox}[colback=white]
   starttime=``0.86'' endtime=``0.98'' word=I \\
   starttime=``0.98'' endtime=``1.1''  word=like \\
   starttime=``1.1'' endtime=``1.40'' word=apples \\
   starttime=``1.45'' endtime=``1.55'' word=but \\
   starttime=``1.55'' endtime=``1.62'' word=not \\
   starttime=``1.62'' endtime=``2.0''  word=bananas\\
    \includegraphics[width=\columnwidth, trim=5cm 22cm 7cm 2cm, clip]{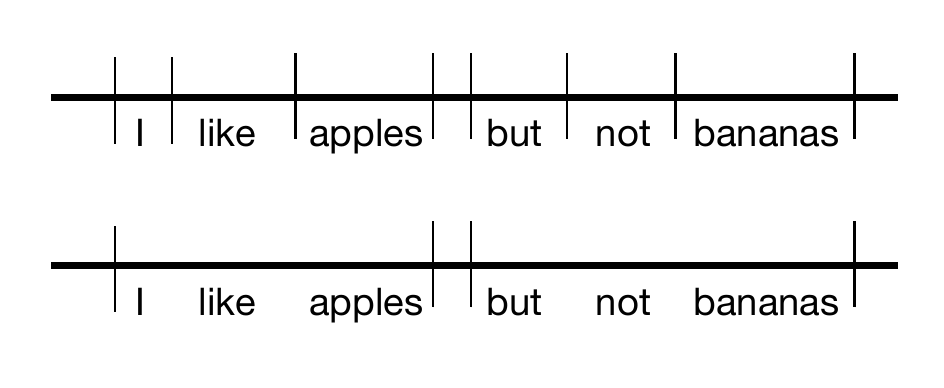}
\end{tcolorbox}

    If adjacent speech segments from the same speaker are not merged, it truly affects the diarization evaluation when collars are considered, as the collar is applied over all VAD borders and these ``break points'' are considered as one of these borders. 
    These ``break points'' between adjacent speech segments are common in the references derived with the Kaldi recipe.
    With our processing, the above transcription results in one speech segment from 0.86 to 1.40 (merged ``I like apples'') and another from 1.45 to 2.0 (merged ``but not bananas'').

    On the other hand, consecutive speech segments from the same speaker separated by pauses (silence) are not merged in any case. Using the above example again, we could think that the pause is too short and maybe it should be discarded and the two segments should be merged into one. But this kind of processing would require some heuristic to determine which is the required pause for considering separate speech segments. We prefer to follow a clean approach keeping the original pauses.

    Anyway, in the case of using no collar (which we believe is the best choice for evaluation) merging the adjacent speech segments has no effect at all. Still, we establish it in case anyone would prefer to use a collar.
\end{itemize}

In later experiments, we use the same partition and references to evaluate our system with two audio types: AMI Mix-Headset audios and the beamformed microphone array N1, where BeamformIt \cite{Anguera07Beamforming}  
is applied using the specific setup provided for AMI. 

The partition, references and audios are shared in our repository\footnote{\url{https://github.com/BUTSpeechFIT/AMI-diarization-setup}}.
This evaluation protocol will also be adopted in the latest Kaldi and pyannote recipes. 
Additionally, we also generated an extra version of the references including all (time-labeled) vocal sounds. As mentioned before, we consider these references not well defined, but we understand that some researchers might find them useful.

\section{Results}

\subsection{CALLHOME}

We first present in Table~\ref{tab:callhome-scoring} the results of our diarization model on CALLHOME data. 
As a baseline, we also provide the result of a standalone AHC clustering of x-vectors, where its threshold is tuned for optimal performance.
We provide these results to illustrate the gain achieved specifically by VBx diarization, which uses AHC as initialization (see section \ref{diarizationsetup} for details).
Our VBx system achieves 4.42\% DER on the \textit{forgiving} evaluation setup, outperforming all systems from previously published works. 
With the \textit{fair} evaluation setup, considering also the overlapped speech (which our system does not handle), performance  drops to 14.21\% DER. Note, that the result achieved in \cite{horiguchi2020endtoend} is obtained on a subset of the CALLHOME dataset, which makes it not directly comparable with our results. The script provided with their implementation makes random partitions of CALLHOME dataset into dev/eval so it is not possible to replicate this partition to make a fair comparison.
Finally, with zero collar (\textit{full} evaluation setup) increases the error to a total 21.77\% DER. We believe that future research works should report results using this more challenging evaluation setup. 

\begin{table*}[htb]
    \centering
    \caption{\label{tab:callhome-scoring} \textit{Diarization performance on CALLHOME. Results marked with * are obtained on a subset of the dataset and are therefore not comparable, see text for more details.}}
	\setlength\tabcolsep{4pt}
    \begin{tabular}{cccl|c|c|c|} 
        \toprule
        \multicolumn{3}{c}{Evaluation setup} &  \multirow{2}{*}{System} &  \multirow{2}{*}{SER} &  \multirow{2}{*}{DER} &  \multirow{2}{*}{JER} \\
        Name & Collar & Overlap &  &  &  &  \\
         \midrule
        \multirow{7}{*}{\textit{Forgiving}} & \multirow{7}{*}{0.25} & \multirow{7}{*}{No} & Kaldi (Sell et al. \cite{Sell2018b}) & \multicolumn{2}{c|}{6.48} & --\\
        & & & Zhang et al.~\cite{zhang2019fully} & \multicolumn{2}{c|}{7.60} & --\\
        & & & Lin et al.~\cite{lin2019lstm} & \multicolumn{2}{c|}{6.63}  & -- \\
        & & & Pal et al. \cite{pal2020metalearning} & \multicolumn{2}{c|}{6.76} & -- \\
        & & & Aronowitz et al.~\cite{Aronowitz2020New} & \multicolumn{2}{c|}{5.10} & -- \\
        & & & AHC  & \multicolumn{2}{c|}{8.10} & -- \\
        & & & VBx  & \multicolumn{2}{c|}{\textbf{4.42}} & -- \\
        \midrule
        \multirow{3}{*}{\textit{Fair}} &  \multirow{3}{*}{0.25} &  \multirow{3}{*}{Yes} & Horiguchi et al. \cite{horiguchi2020endtoend} & -- & 15.29* & -- \\
        & & & AHC & 7.53 & 17.64 & -- \\
        & & & VBx & 4.10 & 14.21 & -- \\
        \midrule
        \multirow{2}{*}{\textit{Full}} &  \multirow{2}{*}{0} & \multirow{2}{*}{Yes} & AHC & 11.06 & 25.61 & 35.48 \\
        & & & VBx & 7.22 & 21.77 & 34.02 \\
        \bottomrule  
   \end{tabular}
\end{table*}

\subsection{AMI}

We first report in Table~\ref{crazyAMI} results for our system using the different evaluation protocols found in the literature for AMI corpus. 
As explained in section~\ref{sec:amiliterature}, the mix of protocols between sites called for running experiments with 3 different data partitions (train/dev/eval sets), 5 sets of references, 2 different types of audio and considering different evaluation setups to be able to compare with all works.
We would like to highlight again that the different protocols differ largely in the amount of speech used for evaluation: from only 13309 seconds of speech considered in \cite{pal2020metalearning} up to 52317 seconds considered in \cite{Bredin2020}.
\begin{sidewaystable}
    \centering
    \caption{\label{tab:ami-pyannote-scoring} \textit{Diarization performance of the proposed model compared to baselines from the literature. (*) denotes  dev and eval pooled results} }
    \label{crazyAMI}
	\setlength\tabcolsep{4pt}
    \begin{tabular}{c|c|c| ccc|c|l|c|c|c|c|} 
        \toprule
        \multirow{2}{*}{Partition} & \multirow{2}{*}{References} & \multirow{2}{*}{Audio type} & \multicolumn{3}{c|}{Evaluation setup} & \multirow{2}{*}{\begin{tabular}{@{}c@{}}Scored  speech \\ dev/eval (s)\end{tabular}} & \multirow{2}{*}{System} & \multicolumn{2}{c|}{development} & \multicolumn{2}{c|}{evaluation} \\
        & & & Name & Collar & Overlap & & & SER & DER & SER & DER \\
        \midrule
        \multirow{5}{*}{\textit{Pyannote}} & \multirow{5}{*}{\textit{Pyannote}} &
        \multirow{5}{*}{\textit{Mix-Headset}} & \multirow{2}{*}{\textit{Forgiving}} & \multirow{2}{*}{0.25} & \multirow{2}{*}{No} & \multirow{2}{*}{29200/29609} & Bredin et al.\cite{Bredin2020} & \multicolumn{2}{c|}{--} &  \multicolumn{2}{c|}{4.6}   \\
        & & & & & & & VBx & \multicolumn{2}{c|}{2.14} & \multicolumn{2}{c|}{2.17}  \\
        \cline{4-12}
        & & & \multirow{3}{*}{\textit{Full}} & \multirow{3}{*}{0} & \multirow{3}{*}{Yes} & \multirow{3}{*}{54051/52317} & Bredin et al. \cite{Bredin2020} & -- & --  & -- & 24.8  \\
        & & & & & & & Bullock et al. \cite{bullock2019overlapaware} & -- & --  & 7.2 & 23.8  \\
        & & & & & & & VBx & 3.33 & 22.95  & 3.86 & 22.85  \\
        \midrule
        \midrule
        \multirow{11}{*}{\textit{Kaldi}} & \multirow{4}{*}{\begin{tabular}{@{}c@{}}\textit{Force}\\\textit{Aligned}\end{tabular}} & \multirow{4}{*}{\begin{tabular}{@{}c@{}}\textit{Beamformed}\\\textit{mic-array}\end{tabular}} & \multirow{2}{*}{\textit{Forgiving}} & \multirow{2}{*}{0.25} & \multirow{2}{*}{No} & \multirow{2}{*}{15053/14080} & Sun et al. \cite{sun2020combinationCambJour} & \multicolumn{2}{c|}{16.4}  & \multicolumn{2}{c|}{15.4}  \\
        & & & & & &  & VBx & \multicolumn{2}{c|}{1.32} & \multicolumn{2}{c|}{1.84} \\
        \cline{4-12}
        & & & \multirow{2}{*}{\textit{Fair}} & \multirow{2}{*}{0.25} & \multirow{2}{*}{Yes} & \multirow{2}{*}{16241/14886} & Sun et al. \cite{sun2020combinationCambJour} & -- & 19.4  & -- & 17.8  \\
        & & & & & & & VBx & 1.26 & 4.96  & 1.92 & 4.67 \\
        \cline{2-12}
         & \multirow{2}{*}{\textit{Kaldi}} & \multirow{2}{*}{\textit{Mix-Headset}} & \multirow{2}{*}{\textit{Forgiving}} & \multirow{2}{*}{0.25} & \multirow{2}{*}{No} & \multirow{2}{*}{18743/18219}  
         & Maciejewski et al. \cite{Maciejewski2018AMI} & \multicolumn{2}{c|}{--}  & \multicolumn{2}{c|}{
        -- / (4.8*)}  \\
        & & & & & & & VBx & \multicolumn{2}{c|}{2.14}  & \multicolumn{2}{c|}{3.02/(2.58*)} \\
        \cline{2-12}
         & \multirow{5}{*}{\textit{Pyannote}} & \multirow{5}{*}{\textit{Mix-Headset}} &\multirow{3}{*}{\textit{Full}} & \multirow{3}{*}{0} & \multirow{3}{*}{Yes} & \multirow{3}{*}{35495/33953} &   Raj et al. \cite{raj2020multiclass} & -- & --  & 10.1 & 23.6  \\
        & & & & & & &   Raj et al. \cite{raj2020doverlap}  & -- & 21.6  & -- & 20.5  \\
        & & & & & & & VBx & 3.12 & 22.63 & 3.56 & 23.47  \\
        \cline{4-12}
        & & & \multirow{2}{*}{\textit{--}} & \multirow{2}{*}{0} & \multirow{2}{*}{No} & \multirow{2}{*}{22812/21911} &   Raj et al. \cite{raj2020doverlap} & \multicolumn{2}{c|}{7.7}  &\multicolumn{2}{c|}{5.2}  \\
        & & & & & & & VBx & \multicolumn{2}{c|}{4.08} & \multicolumn{2}{c|}{3.80}  \\
        \midrule
        \midrule
        \multirow{3}{*}{\begin{tabular}{@{}c@{}}\textit{Kaldi}\\\textit{no TNO}\end{tabular}} & \multirow{3}{*}{\begin{tabular}{@{}c@{}}\textit{Work}\\\textit{specific}\end{tabular}} & \multirow{3}{*}{\begin{tabular}{@{}c@{}}\textit{Beamformed}\\\textit{mic-array}\end{tabular}} & \multirow{3}{*}{\textit{Forgiving}} & \multirow{3}{*}{0.25} & \multirow{3}{*}{No} & \multirow{3}{*}{14545/13309} & \multirow{2}{*}{Pal et al. \cite{pal2020metalearning}} & \multicolumn{2}{c|}{5.02} & \multicolumn{2}{c|}{4.92} \\ 
        & & & & & & & & \multicolumn{2}{c|}{6.21}  & \multicolumn{2}{c|}{2.87} \\
        & & & & & & & VBx & \multicolumn{2}{c|}{4.27} & \multicolumn{2}{c|}{4.58} \\
        \bottomrule 
   \end{tabular}
\end{sidewaystable}
When analyzing the results, it can be seen that our VBx system attains the best results on all evaluation protocols, with two exceptions. 
The first exception is \cite{raj2020doverlap}, which uses Kaldi partition, Pyannote references and \textit{Full} evaluation setup.
This work presents a fusion of 3 different diarization systems dealing with overlapped speech, as compared to our single system with no overlap handling. 
The same system evaluated only on non overlap regions obtains 7.7\% and 5.2\% DER on dev and eval respectively, while VBx obtains only 4.08\% on dev and 3.8\% on eval.
The second exception is the eval result from \cite{pal2020metalearning}, which uses Kaldi ``no TNO'' partition. While this system has significantly better performance than our system (only) on the eval set (2.87 vs. 4.58 DER), when analyzing the results from the paper, it seems to us that this number is an outlier inconsistent with the other results presented in the work: all systems presented in the paper have consistently similar performance on dev and eval sets, this particular system reduces the error more than a 50\% (6.21\% DER on dev vs 2.87\% on eval). Also, this would not be the system of choice if selected according to the performance on the dev set, for which the system was tuned (also shown in the table, with 5.02\% DER on the dev set and 4.92\% DER on eval). Out of all the partitions used, this is the one with the smallest eval set, which might result in noisier results.

In Table~\ref{crazyAMI}, we have demonstrated that VBx method has superior performance as compared to other published works on AMI dataset. However, to offer a fair comparison, we had to deal with too many protocols.
This led us to the proposal of the new evaluation protocol, described in section \ref{newAMIprotocol} which is, already being adopted by other research labs. 
In Table~\ref{tab:ourAMI}, we report results obtained with the VBx system with the proposed evaluation protocol.
This results can serve as a reference for future works on this corpus.
Once again, results are also provided for the baseline standalone AHC when it is tuned for optimal performance (see section \ref{diarizationsetup} for details).
For the sake of completeness, we provide results of our system when evaluated on beamformed mic-array audios as well as on Mix-Headset audios. The system is evaluated on all evaluation setups presented in \ref{sec:eval}. 
Performance on Beamformed audios gets as low as 3.9\% DER on the eval set when using the forgiving evaluation. When considering overlapped speech, DER increases to 14.23\%. Finally, without any collar, the system achieves 20.84\% DER and 26.92\% JER.
For the Mix-Headset audios, results are consistently better, as expected, achieving 2.10\%, 12.53\% and 18.99\% DER for forgiving, fair and full evaluation setups, respectively.

\begin{table*}[htb]
    \centering
    \footnotesize
    \caption{\label{tab:ami-Cambridge-scoring} \textit{Diarization performance of the proposed model on AMI with the proposed AMI protocol.}}
	\setlength\tabcolsep{4pt}
	\label{tab:ourAMI}
    \begin{tabular}{ccccl|c|c|c|c|c|c|} 
        \toprule
        \multirow{2}{*}{\begin{tabular}{@{}c@{}}Audio\\type\end{tabular}} & \multicolumn{3}{c}{Evaluation setup} & \multirow{2}{*}{System} & \multicolumn{3}{c|}{development} & \multicolumn{3}{c|}{evaluation} \\
        & Name & Collar & Overlap & & SER & DER & JER & SER & DER & JER  \\
        \midrule
        \multirow{6}{*}{\rotatebox[origin=c]{90}{Beamformed}}& \multirow{2}{*}{\textit{Forgiving}} & \multirow{2}{*}{0.25} & \multirow{2}{*}{No} & AHC & \multicolumn{2}{c|}{6.32}  & -- & \multicolumn{2}{c|}{7.65}  & -- \\
        & & & & VBx & \multicolumn{2}{c|}{2.80} & -- & \multicolumn{2}{c|}{3.90} & -- \\
        \cline{2-11}
        & \multirow{2}{*}{\textit{Fair}} & \multirow{2}{*}{0.25} & \multirow{2}{*}{Yes} & AHC & 6.43 & 14.68 & -- & 8.82 & 18.36 & -- \\
        & & & & VBx & 2.57 & 10.81 & -- & 4.69 & 14.23 & --\\
        \cline{2-11}
        & \multirow{2}{*}{\textit{Full}} & \multirow{2}{*}{0} & \multirow{2}{*}{Yes} & AHC & 8.68 & 22.14 & 25.29 & 10.93 & 25.48 & 29.85 \\
        & &  & & VBx & 4.20 & 17.66 & 22.26 & 6.28 & 20.84 & 26.92 \\
        \midrule
        \multirow{6}{*}{\rotatebox[origin=c]{90}{Mix-Headset}}& \multirow{2}{*}{\textit{Forgiving}} & \multirow{2}{*}{0.25} & \multirow{2}{*}{No} & AHC & \multicolumn{2}{c|}{3.90} & -- & \multicolumn{2}{c|}{3.96} & -- \\
        & & & & VBx & \multicolumn{2}{c|}{1.56} & -- & \multicolumn{2}{c|}{2.10} & -- \\
        \cline{2-11}
        & \multirow{2}{*}{\textit{Fair}} & \multirow{2}{*}{0.25} & \multirow{2}{*}{Yes} & AHC & 4.06 & 12.31 & -- & 5.05 & 14.60 & -- \\
        & & & & VBx & 1.43 & 9.68 & -- & 2.98 & 12.53 & --\\
        \cline{2-11}
        & \multirow{2}{*}{\textit{Full}} & \multirow{2}{*}{0} & \multirow{2}{*}{Yes} & AHC & 6.16 & 19.61 & 23.90 & 6.87 & 21.43 & 25.50 \\
        & & & & VBx & 2.88 & 16.33 & 20.57 & 4.43 & 18.99 & 24.57 \\
        \bottomrule
   \end{tabular}
\end{table*}

\subsection{DIHARD II}

\begin{sidewaystable}
    \centering
    \caption{\label{tab:dihard-scoring} \textit{Diarization performance of the proposed model on DIHARD II. (*) denotes our own previous results. Results in brackets were obtained when not using the dev set for training nor adaptation purposes, thus they are comparable to our results.}}
	\setlength\tabcolsep{4pt}
    \begin{tabular}{ccclccc|ccc} 
        \toprule
        \multicolumn{3}{c}{Evaluation setup} & \multirow{2}{*}{System} & \multicolumn{3}{c|}{development} & \multicolumn{3}{c}{evaluation} \\
        Name & Collar & Overlap & & SER & DER & JER & SER & DER & JER \\
        \midrule
        \multirow{5}{*}{\textit{Full}} & \multirow{5}{*}{0} & \multirow{5}{*}{Yes} & Landini et al.* \cite{landini2020but} & -- & 17.90 (18.34) & -- & -- &  18.21 (19.14) & -- \\
        & & & Lin et.al. \cite{Lin2020dihard} & -- & 21.36 & -- & -- & 18.84 & -- \\
        & & & Lin et.al. \cite{Lin2020selfattentive}& -- & 18.76 & -- & -- & 18.44 (19.46) & -- \\
        & & & AHC & 10.89 & 21.68 & 42.28 & 13.89 & 23.59 & 43.93 \\
        & & & VBx & 7.41 & 18.19 & 42.53 & 8.85 & 18.55 & 43.91 \\
        \midrule
        \multirow{2}{*}{\textit{Fair}} & \multirow{2}{*}{0.25} & \multirow{2}{*}{Yes} & AHC & 8.22 & 14.91 & -- & 10.94 & 16.67 & -- \\
        & & & VBx & 5.53 & 12.23 & -- & 6.55 & 12.29 & -- \\
        \bottomrule  
   \end{tabular}
\end{sidewaystable}

Diarization results obtained on the DIHARDII dataset are presented in Table~\ref{tab:dihard-scoring}. 
Our VBx system obtains 18.19\% DER on the development set and 18.55\% DER on the evaluation set. 
When analyzing the table, it can be seen that, in fact, these are not the best numbers ever published on DIHARDII eval as the system does not overcome the results from \cite{Lin2020selfattentive}, nor the ones we attained on the challenge \cite{landini2020but} which are, as far as we know, still the best in the literature.
However, note that the best results on DIHARDII \cite{landini2020but} are obtained with  the same VBx method as described in this paper, but including additional adaptation to the DIHARD data and additional steps (overlapped speech handling and resegmentation).
Similarly, in \cite{Lin2020selfattentive} the diarization system is adapted to the DIHARD dev data.
As mentioned in section \ref{diarizationsetup}, in this paper we are aiming to use a generic approach to all datasets, without using the development sets for training or adaptation purposes. 
If we compare our system to the results of \cite{landini2020but,Lin2020selfattentive} when not performing adaptation on the dev set, which are shown in brackets, our current VBx outperforms both systems.

When evaluating the system with the \textit{fair} setup, DER improves around a 30\% on both sets, resulting in 11.75\% and 12.41\%  on dev and eval, respectively.


\section{Conclusion}
\label{sec:Conlusion}

This paper presents a diarization system based on a Bayesian HMM model for clustering x-vectors, also known as VBx.
Our VBx diarization achieves state-of-the-art results on CALLHOME, AMI and DIHARDII without performing specific model adaptation to any of the datasets.

Most of the papers,  which we have compared our system with, present clustering of new sets of embeddings/x-vectors. All these approaches are complementary with VBx: the VBx clustering can be combined with most of these new embeddings, which shows the further potential of the method.

In the case of AMI dataset, we have presented, for the first time, a fully fair comparison of our system with several works from the literature.
One of the major contributions of the paper is the proposed evaluation protocol for AMI, which will be adopted also in future Kaldi and Pyannote recipes, and which we hope will become a new standard.

The analysis of results for CALLHOME and AMI datasets reveals that systems are reaching very low diarization error rates when evaluating with the standard 0.25\,s collar and without considering overlapped speech regions. We believe that, as in latest diarization challenges, systems should be tested using more challenging evaluation setups considering also overlapped speech and no collar.

Future work will focus on combining our system with real VAD instead of the oracle labels, to approach the setup of nowadays largely popular end-to-end techniques.


\section{Acknowledgements}
\label{sec:Ack}
We would like to thank Brian Sun \cite{sun2020combinationCambJour,Fathullah20}, Monisankha Pal \cite{pal2020metalearning}, Matthew Maciejewski \cite{Maciejewski2018AMI} and Desh Raj \cite{raj2020multiclass,raj2020doverlap} for detailed feedback about their evaluation protocols for AMI. 
Special thanks to Desh Raj, Paola Liebny García-Perera and Herv\'e Bredin for their contribution on the new evaluation protocol for AMI. 

\bibliography{biblio}

\end{document}